\documentstyle [12pt,aasms4]{article}

\newcommand{\mo}{ \cal  M_{\odot}  }

\newcommand{\amin}{^{\prime}}                     
\newcommand{\asec} {^{\prime\prime}}                     
\newcommand{\adeg} {^{\circ}}                     
\newcommand{\rasec}{\hbox{$\,$\raise 0.6 ex \hbox{\rm s}\kern-.35em
                  \lower 0.0 ex \hbox{.}$\,$}}        
\newcommand{\decsec}{\hbox{$\,$\raise 0.0 ex \hbox{$\asec$}\kern-.45em
                  \lower 0.0 ex \hbox{.}$\,$}}         
\newcommand{\decmin}{\hbox{$\,$\raise 0.0 ex \hbox{$\amin$}\kern-.45em
                  \lower 0.0 ex \hbox{.}$\,$}}         
     
\newcommand{\gtabouteq}{\,\hbox{\raise 0.5 ex \hbox{$>$}\kern-.77em 
                    \lower 0.5 ex \hbox{$\sim$}$\,$}}       
\newcommand{\ltabouteq}{\,\hbox{\raise 0.5 ex \hbox{$<$}\kern-.77em 
                     \lower 0.5 ex \hbox{$\sim$}$\,$}}  
\begin{document}


\slugcomment{Submitted to A.J.}

\lefthead{Irwin, English, and Sorathia}
\righthead{Radio Halos}

\title{High Latitude Radio Emission in a Sample of Edge-On Spiral Galaxies}

\author{Judith A. Irwin}
\affil{irwin@astro.queensu.ca}

\author{Jayanne English}
\affil{english@astro.queensu.ca}

\and

\author{Barkat Sorathia\altaffilmark{1}}
\affil{sorathia@polaris.phys.yorku.ca}

\vskip 0.5truein

\affil{Queen's University}
\centerline{Dept. of Physics}
\centerline{Kingston, Ontario, Canada, K7L 3N6} 
\vskip 0.5truein


\altaffiltext{1}{Now at York University, Physics \& Astronomy Dept.,
4700 Keele Street, North York, Ontario, Canada, M3J 1P3}


\begin{abstract}
We have mapped
 16 edge-on galaxies at 20 cm
using the VLA in its C configuration,
and a subset of these galaxies in D configuration at 6 and/or 20 cm
in a search for extended ($\gtabouteq$ 1 kpc)
radio continuum emission above and below
the plane.  For 5 galaxies, we could form spectral index, energy
and magnetic field maps (assuming minimum energy).
While the galaxies were partly chosen by radio flux density, 
they span a variety
of star formation rates and only 6 might be considered ``starburst"
galaxies.   A range of Hubble type and degree of isolation are also
represented.  The galaxies largely fall on the FIR-radio continuum
correlation.  They also display a correlation between IR surface
brightness and warmth, extending the previously observed relation of
Lehnert \& Heckman~(1996b)
to galaxies with lower star formation rates.

We find that all but one galaxy show evidence for 
non-thermal
high latitude radio continuum emission, suggesting that cosmic ray
halos are common in star forming galaxies.
  Of these,  eight
galaxies are new detections.  
The high latitude emission is
seen over a variety of spatial scales and in discrete and/or smooth
features.  In some cases, discrete features are seen on large scales,
suggesting that smooth radio halos may consist, in part, of
discrete features combined
with low spatial resolution. 
 In general, the discrete features emanate from
the disk, but estimates of CR diffusion lengths suggest that diffusion
alone is insufficient to transport the particles to the high latitudes
seen ($>$ 15 kpc in one case).  Thus CRs likely
diffuse through low density
regions and/or are assisted by other mechanisms (e.g. winds).

We searched for correlations between the prevalence of
high latitude radio emission and
a number of
other properties, including the global SFR, 
supernova input rate per unit star forming
area, \.E$\rm_A$, and environment, and do not find clear correlations
with any of these properties. A subset of the data allow, at best, for only
a weak correlation with \.E$\rm_A$.
Our one non-detection (NGC~4517), however, occurs at a
threshold level consistent with that found by Dahlem et al. (1995)
The lack of a good correlation with star formation
indicators could be the result of the different timescales for
star formation processes compared to the duration of the radio emission. 
 Correlations with
other properties, such as environment, are more difficult to assess.
However, a few isolated galaxies display strong radio halos, indicating
that an interaction is not necessary to produce the
extra-planar emission.

\end{abstract}


\section{Introduction}
\label{intro}

There is now considerable evidence for galactic halos, thick disks, 
and discrete connecting features between the disk and halo in many
 star-forming spiral galaxies.  Such features are best
detected directly by observing edge-on galaxies with sufficient
sensitivity and spatial
resolution.  High latitude (i.e. $\gtabouteq$ 1 kpc)
emission has now been
observed in every ISM tracer (though not necessarily in all tracers
in the same galaxy)
for example,
CO (\cite{is96}),
IR emission from dust (\cite{hs97}),
HI  (\cite{li97}), 
diffuse ionized gas (Rand et al. 1990; Dettmar 1990),
X-ray emission from the hot ionized medium (\cite{vp96}),
and radio synchrotron emission from cosmic rays (CRs) 
(\cite{enir97}).  For a more complete list, see \nocite{dahl97} Dahlem
(1997).

To account for emission in galactic halos, several models have been
 developed which include
some form of gas ejection from the disk to the halo, with subsequent
infall (circulation or ``fountains") or ejection into the intergalactic
medium (\cite{sf76}; \cite{bre80}; \cite{ni89}; \cite{ham90};\cite{vol91}).  
Processes related to massive star formation (e.g. supernovae, stellar
winds) are the focus of these models, although magnetic fields through
Parker instabilities 
(\cite{ka96}), radiation pressure on dust (\cite{fr91})
and galaxy-galaxy interactions (\cite{gh94})
have also been suggested as playing
an important role in some cases.
With estimated disk mass-loss rates of 0.3 $\to$ 3 M$_\odot$ yr$^{-1}$
for our own Galaxy (\cite{ni89})
and most likely
 higher for starbursts (see \cite{ham90}; \cite{lh96s}; \cite{lh96})
there are important
implications for metallicity gradients, 
the availability of disk gas in an appropriate state for star formation, 
and
 indeed for galaxy evolution itself.  Faint extended halos, when present, may
also trace the more arcane dark matter distributions around
galaxies.

Early evidence for high latitude radio features in
external spiral galaxies was presented some years ago
 by \nocite{hvk83} Hummel, van Gorkom, \& Kotanyi (1983) who identified 5
edge-on galaxies with discrete radio continuum features 
apparently associated with nuclear outflow.  This prompted one of
us (JAI) to initiate a
small survey to search for evidence
for high latitude radio continuum emission in other edge-on galaxies, 
whether connected with the underlying disk or nucleus.
Since the time of the initial survey, various follow-up observations
and radio continuum maps of a few individual galaxies 
in the sample have been
published (see \cite{irw91}; \cite{irw95}; 
\cite{enir97};  \cite{li97}; 
\cite{ki97} 
\cite{dib98}).  In this paper, we present the complete survey results
(see also \cite{sor94}). 
For recent reviews of gaseous halos of
spiral galaxies see \nocite{det92} Dettmar (1992) and \nocite{dahl97}
Dahlem (1997).

\section{The Sample} 
\label{sample}

The edge-on galaxies in this sample were
chosen from the 1415 MHz Westerbork survey of 
\nocite{hum80} Hummel
 (1980; 190 galaxies), 
supplemented by the 2380 MHz Arecibo survey of \nocite{drecon78}
Dressel \&
Condon (1978; 456 galaxies) with declinations high enough for
observations at the Very Large Array.
A galaxy was considered to be edge-on if its
inclination, i$\ge$ 72 $\adeg$ (log R$_{25}$ $\ge$
0.52 as listed in the \nocite{RC2}
de Vaucouleurs et al. 1976; hereafter the RC2).
We also restricted the sample to galaxies whose total flux densities
at 1415 MHz were greater than 50 mJy  ($>$ 30 mJy at 2380 MHz)
to ensure that they could be
mapped in reasonable integration times. 
Thus, the sample is
 biased towards galaxies with
higher radio flux density, but as we will see in 
\ref{derive}, this does not translate
into a bias towards high star formation rate (SFR).
A few galaxies of large angular extent were also excluded 
so as to avoid cases in which
halo emission would likely be resolved out in the first data set
acquired and also to avoid galaxies which had already been
well studied (e.g. M~82).  
In total, 16 galaxies met the above criteria.  These galaxies are listed,
along with some of their properties, in Table~\ref{galcharacter}  
(we adopt $\rm H_o = 75 \ km \ s^{-1} \ Mpc ^{-1}$).

The most extensive radio continuum survey of edge-on spirals carried
out to date is that of \cite{hbd91} who surveyed 181
galaxies with inclinations $>$ 75$\adeg$ at a frequency of
5 GHz.  Most of their galaxies
were restricted to optical diameters
$<$ 4$^\prime$ and therefore there is only one galaxy, NGC~5775
(UGC~9579), which
is common to our respective samples.  Hummel et al. detect radio halos in
only 5.5\% of their galaxies.
 The main differences between the two
surveys are  a) we have looked at a smaller radio-selected sample,
 b) our galaxies are generally of larger angular size and
thus we have improved linear resolution, c) with roughly equivalent
beam sizes, our sensitivity ranges from a factor of 2 better to a value
comparable to that of Hummel et al. (cf. our Table~\ref{beams} with
their Table 3) and d) our observations are at 1.49 GHz, supplemented with
some 5 GHz observations (\S\ref{obsred}), compared to Hummel et al.'s 
5 GHz data.

\cite{col96} have also observed 10 edge-on Seyfert galaxies
and 4 non-Seyferts at a frequency of 5 GHz
and find that 6 out of 10 Seyferts display extended 
 emission apparently originating at the nucleus.  
Two of their galaxies, NGC~4388 (a Seyfert) and
NGC~3044 are in our sample.

\section{Observations and Data Reductions} 
\label{obsred}
The first observations were made on
25, 26 July, 1985 at 20 cm (1.49 GHz) using the Very Large Array (VLA)
 of the National Radio Astronomy Observatory (NRAO)
\footnote{\small The NRAO is operated by Associated Universities
Inc., under contract with the National Science Foundation, USA.}
in its C configuration.
Subsequently, on 27, 28 April 1987 with the VLA in its
D configuration, 11 galaxies from the sample
 were observed at 20 cm and 5 galaxies were observed at 6 cm (5 GHz).
For each observation, the bandwidth was 100 MHz, the
primary flux calibrators were 3C48 or 3C286 and nearby phase
calibrators were observed every 20 to 30 minutes.  The largest
angular size visible is 4 arcmin
for the C array 20 cm and D array 6 cm observations
and 15 arcmin for the D array 20 cm observations. 
The half-power primary beam size is 30 arcmin at 20 cm and
9 arcmin at 6 cm.  On-source
integration times were chosen
according to prior knowledge of the galaxy's total flux and
ranged from a minimum of 13 minutes to a maximum of
2.8 hours over
all observations.
The mean on-source time for the C array 20 cm data was 48 minutes,
 for the
D array 20 cm data, the mean was 1.3 hours, and for the
D array 6 cm data, the mean was 58 minutes.
  
For historical reasons, related in part to the way in which the
data have been handled over a number of years, we have dealt with
each array separately in presenting an overview of
the complete survey results.
Combined array images have been presented for several individual galaxies
elsewhere (see \cite{bdi93};  \cite{dib98}).
For each array, then, the UV data  
were Fourier Transformed into the map plane using
both uniform and natural weighting,
CLEANed, and self-calibrated (see Perley et al. 1989).
The resulting 
rms noise and synthesized beam parameters
are listed in Table~\ref{beams}. 
The highest linear resolution (cf Columns 2, 4, and 5) 
ranges from 0.6 to 6.4 kpc, depending on distance.
  In general, the rms noise
is approximately twice the theoretical value for the 20 cm 
observations (due to background confusing sources and their
sidelobes) and is approximately equal to the theoretical
value for the 6 cm observations.  All maps were then corrected
for primary beam attenuation.  
 
\section {Results}
\label{results}

The intensity maps are displayed in Appendix~A. 
 The first is the uniformly weighted C array
20 cm map superposed on an  
optical image from the Digitized Sky Survey 
(DSS)\footnote{\small DSS images were
acquired via the SkyView website developed and maintained under
the NASA ADP Grant NAS5-32068.}, the second is
the naturally weighted C array 20 cm map showing the optical major
and minor axes, the third is the naturally weighted D array 20 cm
map also showing the optical axes 
 and the fourth is the naturally weighted D array
6 cm map (the latter two, if they exist, cf. Table~\ref{beams}).
Following these are a series of derived maps, i.e. spectral
index, total energy, and magnetic field (\S\ref{minen}) if they exist.
Contour levels and/or greyscale ranges
for all displayed maps are given in Tables A1 to A5.
Note that all displayed intensity maps
are uncorrected for the primary beam, but all derived maps (e.g.
spectral index, etc.) and other derived quantities include
the primary beam correction.

\subsection{Extra-Planar Radio Continuum Emission}

\subsubsection{Optical and Radio Dimensions of the Galaxies}
\label{measdim}

In Table~\ref{radaxes} we present a comparison between the optical
and radio dimensions of the galaxies.  The radio dimensions are strongly
dependent on the array and the weighting.  This
is also true of the semi-minor to semi-major axis ratio, b/a, because
convolution with 
 a large beam selectively increases
 b with respect to a.
Thus, for the radio dimensions
of Table~\ref{radaxes}, we have chosen the 
C array uniformly
weighted data which has the highest spatial resolution 
and for which the minor axis emission is well resolved (for example,
there are at least 2.7 and up 
to 9 beams across the radio minor axes with an average
of 5.3 for the galaxy sample). 
 We then used a deconvolution algorithm
to spatially deconvolve the known
 beam from the emission
over the entire galaxy before listing the linear dimensions
(Columns 4, 5).  All measurements are to the 2$\sigma$ level.
With this correction, the ratio, radio b/a
(Column 7), is 11\% (on average) {\it lower} 
than it would be by taking the apparent radio
minor to major axis ratio without the correction.

The median value of b/a for the optical 
measurements is 0.23 with a standard deviation of  0.04, while
the median radio ellipticity is 0.38 $\pm$ 0.10.  The largest differences
are for galaxies NGC~4746, NGC~4157, and NGC~3044 whose radio b/a range
from 1.7 to 4 times the optical b/a.
In only two cases (NGC~2613 and NGC 4517) is the
radio b/a equivalent to or smaller than the optical value.

\subsubsection{Discrete Radio Extensions}
\label{rad-discrete}

Every edge-on
galaxy (Appendix A)
shows radio spurs or discrete features extending
in a direction away from the major axis
(when each total intensity image is examined),
 with the possible exception
of NGC~5297 and NGC~5433 whose discrete features are 
 marginal.
Also, in a few cases, namely IC~0564, NGC~3735 and NGC~4517, the apparent
extensions are point-like and/or disconnected from the
major axis emission and are therefore discounted as  background sources. 
Thus 11 of 16 galaxies show discrete vertical radio extensions. 
 In all of these
galaxies, at least one or more radio features
extend beyond the optical disk (see
overlay of the first image or subsequent maps showing optical axes)
 indicating that they are 
extra-planar features. 

There are also discrete features which project against the optical disk
of the galaxy and which may or may not be out of the plane. Such 
 features  do not have the appearance
of in-plane structures which usually
trace spiral arms.  For
example, a single radio spur on the north-west side of NGC~2683
(see radio contours over optical
image) is projected against the background optical disk.
If this feature were in the plane, it would delineate a radial
radio feature extending over 3 kpc, isolated from the bulk of the radio
emission interior to it and with no counterpart on
the other side. Since we know of no precedent for such radio
features in the disk and, by analogy with the known 
extra-planar spurs, we also interpret 
smaller discrete features such as the one described
 as extending out of the plane.

The discrete features appear to originate in the galaxy disk from various
positions along the major axis.
 Consequently, searching for high latitude radio continuum emission
by only taking slices along the minor axis would be insufficient 
to detect them.  This also suggests that they originate with disk-related,
rather than nucleus-related, phenomena.   
Possible exceptions in our sample are NGC~3044, which shows a
large spur over the nuclear region and the well-defined and well-known
 nuclear radio 
lobes of NGC~4388 (\cite{hvk83}; \cite{weh87};
 \cite{sto88};
\cite{hs91})
which are visible as 
a bulge in the contours
on either side of the nucleus in the 20cm C array images of Appendix
A.  Large extensions are also visible on the
20 cm D array image, some of which could also be related to the nuclear outflow.
However, the significance of the lower contour levels 
in this particular map is uncertain because of cleaning
difficulties in the environment of other 
radio bright Virgo Cluster galaxies.

Even for these two galaxies in which nuclear features may be present, 
there are also other discrete features which appear to originate in the disk.
See, for example, smaller extensions on the western major axis of
NGC~3044 (C array naturally weighted data).  For NGC~4388, there is
a striking extension to the north, originating in the eastern major
axis as seen in the 6 cm D array data.  Whether this feature extends
as far as implied in this image is uncertain (see above) but the reality
of an extension at this position is not in question because a disturbance
is also seen in the independent 
20 cm C Array image and an extension is also seen at
this position in an HI image kindly supplied by J. van Gorkom.


\subsubsection{Radio Continuum Halos -- Modeling a Thin Disk}
\label{modeldisk}

The D array naturally weighted images (Appendix A) show considerable emission
beyond the projection of the optical minor axis.  However, since
beam smearing should be most pronounced for this lowest resolution
data set, some or all of this emission could simply be due to
beam smearing.  To investigate this,  
 we have generated
a thin disk model galaxy for all cases
for which D array observations exist
(10 galaxies in total, 
Table~\ref{beams}, since the D array map for NGC~1421 was poor) 
and smoothed the inclined thin disk with the known beam to
compare with observations.  A thin disk (and its corresponding minimum
inclination, cf. Column 8, Table~\ref{galcharacter}) is adopted so
as to be conservative and assume that as much emission as possible
is due to the projection of in-plane
emission against the sky.  Thus, if anything, we will underestimate the
extraplanar radio emission.
 We adopted a gaussian intensity distribution
along the major axis and matched the model major axis extent to
 the observed radio continuum
major axis at the 2$\sigma$ level.
 This procedure is similar to that followed by
\nocite{dah97} Dahlem et al. (1997) 
except that we have allowed for a non-uniform 
major axis intensity
distribution and we also model the entire thin-disk distribution,
 rather
than taking only a single slice along the minor axis.
The results are shown in Fig.~\ref{thindisk} in which
the smoothed model  (heavy dashed contour) is plotted over
the observed distribution (thin solid contours).  
Fig.~\ref{thindisk} gives a visual impression of total radio power in
the halo compared to the disk, for a disk conservatively chosen
to contain as much power as possible.

For almost all galaxies, there is some evidence for extended
emission (beyond the smoothed thin disk model) in a direction 
perpendicular to the major axis.  The most obvious examples are
NGC~2683, NGC~3432, NGC~3556, NGC~4388, and NGC~5775.  Galaxies
which show less obvious extensions or halos 
are NGC~2613 (the large south-western feature could be a background
source), NGC~3044, NGC~3221, and NGC~5297.
The remaining galaxy for which D array naturally weighted data
exist, NGC~4517, shows little evidence for extensions with the
possible exception of two weak features (left top) on either
side of a strong background point source. 

Even in these low resolution maps, the extended emission often
has structure.  For example, radio spurs are seen in NGC~3221,
and NGC~5775
and larger discrete extensions are seen in NGC~2683.  The existence of 
structure 
far from the plane has been pointed out before for NGC~3432 by
\nocite{enir97} English and Irwin (1997) and from deeper 
 observations of NGC~5775 (see \cite{irw95}).  Generally,
 as resolution improves, more
discrete structures appear. 

\subsubsection{Prevalence of High Latitude Radio Continuum Emission in
the Galaxies}
\label{radio-rank}

Since the detection of high latitude radio continuum
emission is dependent on sensitivity to various spatial scales,
 we have 
combined the above results
in order to rank the observed  
galaxies 
according to the visibility of their extended radio emission
as follows.  Galaxies with b/a (radio) $>$ f b/a (optical)
are given a weight of 2 for f$\,\ge\,$2, 1 for 1$\,<\,$f$\,<\,$2, and 0 for 
f$\,\le\,$1 (see \S\ref{measdim}).
Galaxies which show discrete extensions beyond the optical disk are given
a weight of 2, marginal cases are given a weight of 1,
and galaxies where there is no evidence for discrete sources or the
features are suspected of being background sources are given a weight
of 0 (outlined in \S\ref{rad-discrete}).
Finally, galaxies with obvious radio halos or extensions beyond
a modeled thin disk (D array natural weight data) are given a weight of
2,
less obvious cases a weight of 1, and no evidence, a weight of 0
(see \S\ref{modeldisk}).
Those galaxies for which
D array data exist are ranked out of a total of 6 and those galaxies
for which D array data do not exist are ranked out of a total of 4.
Using all 3 criteria allows us to take into account evidence for
high latitude radio continuum emission over a variety of spatial scales
from high resolution (C array uniform weight b/a ratio), to intermediate
resolutions (discrete features beyond the optical disk
as illustrated in Appendix A) to low
resolution (D array thin disk modeling).
(Note that we do not attempt to make corrections for
differences in UV coverage.)  Higher values, therefore, indicate
that high latitude radio continuum emission is prevalent over many
spatial scales and/or that the emission is stronger at
any given spatial scale. 

The results are shown in Column 3 of Table~\ref{rank}.   
Only one galaxy, NGC~4517, shows no evidence for extended radio continuum
emission on any scale.
This is also the galaxy in our sample
which has the largest angular size
(10.5 arcmin, cf Table~\ref{galcharacter}); however,
we should have detected high latitude radio continuum emission in this
galaxy, if it
exists, on spatial scales up to 15 arcmin (the largest angular size
detectable at D array 20 cm).  
All other 15 galaxies  show some evidence for
extended high latitude radio continuum emission (a 94\% detection rate),
either a high radio b/a ratio, discrete extensions, 
smooth broadscale structure, or some combination of these.
  For 7 of these galaxies, 
extra-planar radio emission has been observed previously.  These are:
NGC~3044 (\cite{hvh89}; \cite{col96}),
NGC~3221 (\cite{hvh89}),
NGC~3432 (\cite{enir97}; same data),
NGC~3556 (\cite{bdi93}),
NGC~4157 (\cite{hvh89}),
NGC~4388 (nuclear outflow only; \cite{hvk83}; \cite{weh87};
 \cite{sto88};
\cite{hs91}),
NGC~5775 (\cite{hbd91}; \cite{irw95}).
Thus, we have detected extra-planar radio continuum emission 
 in 8 more edge-on galaxies. For two of these 8 galaxies
(NGC~1421 and NGC~2683),
some  high latitude features can be similarly seen in the
maps of \nocite{con87} Condon 1987.

\subsection{Infra-Red and Radio Continuum Parameters}

\subsubsection{Fluxes, Luminosities, and Star Formation Indicators }
\label{derive}

In Table~\ref{luminosity}, we present total flux densities,
luminosities, and other star formation indicators.

The flux density (Columns 2, 3) is measured from the lowest resolution
D array naturally weighted maps, except for galaxies for which
no D array data were available
(cf. Table~\ref{beams}) and 3 galaxies for which the 
C array fluxes were
considered to be more reliable. 
  In the latter category, the galaxies are
NGC~1421 (D array data were poor, possibly due to
the low elevation of the source), NGC~3556 (the C array flux had a
significantly smaller error bar but D and C array flux measurements  
agree to within errors) and NGC~2613 (the C array flux was higher
than the D array flux and closer to previous literature values).  
All of our C array fluxes agree with a previously measured value
(cf. \cite{hum80}; \cite{hphd85}; \cite{con87}; \cite{chss90})
to within error bars or to within 10\% when errors are not
listed. 

Using standard formulae (see \cite{sanmir96}, their Table 1) we calculate
a far infra-red luminosity $\rm  L_{FIR}$ which 
represents emission
in the 
 $\rm 40 \mu \leq \lambda \leq 120 \mu$ range
(uncorrected for longer wavelengths) and which is usually
plotted in the FIR -
radio continuum correlation.  We also calculate the 
infrared luminosity $\rm L_{IR}$ which represents emission over
$\rm 8 \mu \leq \lambda \leq 1000 \mu$ and is an indicator of
luminous starbursts.
The infrared flux densities used in these derivations 
were taken from the 
IR bright ($\rm S_{60\mu} > 5.24$ Jy)
sample catalogued in 
\nocite{sbns89} Soifer et al.~(1989) if available (12 sources) which includes
flux within $\pm$ 4 arcmin of each
source position. 
The values for fainter sources, 
NGC~2613, NGC~4517, NGC~4746, and NGC~5297,  were taken from
the nearest positions in the \nocite{iraspt} 
Infrared Astronomical Satellite (IRAS) 
 Point Source Catalog (PSC)(1988). (The association between the
IRAS source and NGC~2613 is uncertain.)
The latter values include flux only within the IRAS beam
(1 arcmin) which is smaller than the optical extent of the galaxies.
 However, most of the IR luminosity
in galaxies is usually concentrated towards
the central regions and 
we also find
(see Fig.~\ref{FIRradio}) that these galaxies
 fall on the FIR - radio continuum
correlation,  showing no significant discrepancies compared to galaxies
which used the Soifer et al. IR fluxes.  This suggests that
the luminosities for these galaxies agree with those of Soifer et al.
to within $\sim$ 20\%.

For star formation rates (SFRs, Column 9), we follow \nocite{ham90}
Heckman et al. (1990)
and compute this value for
 stars between 0.1 and 100 $\mo$ 
 i.e.
SFR$_{0.1 \to 100 {M_\odot}}$
 = 26  L$_{{IR},11}$ $(M_\odot \,yr^{-1})$,
where L$_{{IR},11}$ is the infrared luminosity
in units of 10$^{11}$ L$_\odot$
(see also \cite{hggrl86}).  This relation assumes that
the bolometric
luminosity, L$_{bol}$, is about equal to 
 L$_{IR}$, an assumption which will underestimate the SFR because it
is clear
(cf. Columns 5 and 6) that there is at least a significant
blue luminosity for these galaxies.  We do not
attempt to include L$_B$ in our estimate of L$_{bol}$
 because of the unknown and
variable internal extinction in these
edge-on systems and also to be consistent with
previous authors. The resulting SFRs can be compared to
  the value for
the canonical starburst galaxy M~82 
of 9 $\rm M_{\odot}\ yr^{-1}$ (\cite{ham90}). 
By this measure, 
6 of the galaxies
in our sample could be considered starburst galaxies.  

Column (10) lists  the distance independent quantity,
\.E$\rm_A$, which represents the
rate of 
 energy input into the ISM by supernovae per 
unit star forming area. 
 This quantity is computed from
\.E$\rm_A =$  $\nu_{SN} \rm E_{SN}$  / $\pi r^2_{SF}$ ergs s$^{-1}$ cm$^{-2}$,
where $\rm E_{SN} = 10^{51}$ ergs is the energy
input per supernova and
$\rm \nu_{SN} = 0.2 \ L_{IR,11}$ SN yr$^{-1}$ 
is the supernova rate (\cite{ham90}).
 The computed  supernova rates
agree with those of 
\nocite{con92} Condon (1992) (who used L$_{FIR}$
rather than L$_{IR}$) to within 5\%.  

For the star forming radius,
 $r_{SF}$,  we use
the highest resolution 20 cm C-array semi-major axis corrected for beam 
smoothing (Table~\ref{radaxes}). 
 The star forming radius is probably
the largest source of error on \.E$\rm_A$ since its measurement
depends on the rms noise in the map  as well as 
CR diffusion distances at the outer disk edge. 
The errors are in opposite directions, the former resulting in an
underestimate of $r_{SF}$ and the latter resulting in an
overestimate of $r_{SF}$.  To investigate the magnitude of these
errors, we
 obtained a very deep
H$\alpha$ image [peak signal-to-noise
(S/N) ratio = 1200/1] of NGC~5775, kindly
supplied by R.-J. Dettmar.  For this galaxy, we find
$r_{SF}$(H$\alpha$) = 13.3 kpc 
which differs from our
radio radius of 12.6 kpc (Table~\ref{radaxes}) by only 5.3\%.
This translates into an 11\% change in \.E$\rm_A$. 
Even though this is small, in the absence of information on
the other galaxies, we have assumed that generous errors of up to 25\% 
in $r_{SF}$
could be present, resulting in error bars on \.E$\rm_A$
of up to 50\%. 

The meaning of \.E$\rm_A $ is the same as \.E$\rm^{tot}_A$ of
\nocite{dlg95} Dahlem, Lisenfeld, \& Golla (1995), except that
they have used the radio continuum power, rather than the
IR luminosity to compute 
$\nu_{SN}$ (for the two galaxies which are common between
the samples, NGC~3044 and NGC~5775, their $\nu_{SN}$ is 2.7 $\times$
higher than ours).  Their measurements of $r_{SF}$, which are also
noise-dependent, are taken from either H$\alpha$
or IRAS CPC images.

\subsubsection{FIR-Radio Continuum Correlation}
\label{fir-rad}

Figure~\ref{FIRradio} plots the distance independent
 FIR-radio continuum correlation for the galaxies in our sample.
The  logarithmic measure of this 
correlation (\cite{helsr85})
is\break
q = log [(L$_{FIR}$/3.75$\times$10$^{12}$ Hz)/P$_{20cm}$), where
 L$_{FIR}$ (W), P$_{20cm}$ (W Hz$^{-1}$) are taken from
Table~\ref{luminosity}.
Excluding the Seyfert 2 galaxies, NGC~3735 and NGC~4388,
we find a mean value for our sample of q = 2.2 and a dispersion
of $\sigma _q$ = 0.1. 
 The latter value is comparable to the
error in measurement.
(Note that there
would be no significant difference to q
if the Seyfert, NGC~3735, were
included since this galaxy also falls on FIR-radio continuum
relation.)
Our estimate of   q is consistent 
with the value of 2.10 ($\sigma _q$ = 0.16) for  FIR 
luminous galaxies (\cite{helsr85})  and 
 2.3 ($\sigma _q$ = 0.2) for normal 
galaxies (\cite{con92} and references therein).  
The discrepant value for NGC~3432 is discussed in 
English \& Irwin (1997). 

The markers in Figure~\ref{FIRradio} indicate the supernova energy
input rate per unit star forming area, from Table~\ref{luminosity}.
There is a weak tendency for IR-luminous galaxies to also
have a higher supernova energy input rate per unit area, but there
are clear exceptions.  Note the difference on the plot between
NGC~4746 and NGC~3221, for example, or between NGC~2683 and NGC~3735
(each pair with similar values of \.E$\rm_A $).

\subsubsection{IR Surface Brightness -- Colour Correlation}
\label{colour-luminosity}

It has been argued that 
a FIR flux ratio  with S$_{60\mu}$/S$_{100\mu} \ > $ 0.4 
indicates that a galaxy may be IR warm (\cite{ham90}).
Figure~\ref{dustenrate} plots this infra-red colour against the supernova 
energy input rate per unit star forming area.
Since 
\.E$\rm_A $ is obtained directly from L$_{IR}$ (\S\ref{derive}), this
plot has the same meaning as
the plot of IR surface brightness versus IR warmth
given by \nocite{lh96}Lehnert \& Heckman~(1996b) for starburst galaxies,
except that we normalize by the radio continuum disk area, rather
than the H$\alpha$ disk. 
 Figure~\ref{dustenrate} 
confirms Lehnert \& Heckman's correlation, and extends their plot to
lower surface brightnesses and cooler colours, i.e. this correlation
also appears to apply to normal, as well as starburst galaxies.
Lehnert \& Heckman suggest that the dominant heating source for dust
is UV radiation from hot stars, as opposed to heating by the general
diffuse radiation field or an active galactic nucleus (AGN).
The fact that we find a similar correlation
using the radio continuum disk size (where the radio continuum 
is presumably correlated with the hot stars)
appears to be consistent with their conclusion.

An exception to this relation is NGC~3432 which has a discordent
SN input rate for its dust temperature.  This difference reflects its
offset position on the radio continuum -- FIR correlation
(Fig.~\ref{FIRradio}), i.e.
its IR luminosity is low in comparison to its radio continuum power.
If \.E$\rm_A $ had been determined from radio continuum power,
 rather than IR luminosity, then this galaxy would have conformed 
to the general trend shown in Fig.~\ref{dustenrate}.

\subsection{Spectral Index, Energy, and Magnetic Field Maps}
\label{minen}

For the 5 galaxies with both 20 cm C array and
 6 cm D array observations (NGC~1421, NGC~3044,
NGC~3556, NGC~4388, and NGC~5775), we could obtain
 spectral index, $\alpha$, maps 
 (S$_\nu\,\propto\,\nu^{\alpha}$), as well as maps of total energy
and magnetic field strength using the minimum energy assumption.
These are shown in Appendix A with contour and greyscale levels
listed in Tables A4 and A5.  Some derived parameters from these maps
are listed in Table ~\ref{enmag}.

\subsubsection{Spectral Index Maps}
\label{specindex}

Spectral index maps (Appendix A,
Table A4) were made from the natural weighting 6 cm D array and 
natural weighting 20 cm C array
 data after smoothing to equivalent beams and applying
a $\rm 1.5\ \sigma$ noise cutoff. 
 The minimum, mean, and maximum values from these maps are listed
in Columns (2) to (4) of Table~\ref{enmag}. Values in parentheses
represent corresponding values as measured from
the uniform weighting spectral index maps which were similarly
formed but are not shown.  

Clearly, the steep spectral indices indicate that the
emission is dominated by non-thermal components
(see also \cite{nik97}).  The 
thermal contribution from these galaxies is unknown; 
however, if we use the guidelines outlined in \nocite{con92} Condon (1992),
then the non-thermal spectral index,
$\alpha_{NT}$, will be 0.1 to 0.3 steeper than the observed spectral
index,  for $\alpha$ ranging from -0.6 to -1.03, respectively 
(Table~\ref{enmag}).  The uncertainty in $\alpha$ is also of order 0.1.
Since the observed variations in $\alpha$ are of magnitude 1 to 2 
(Table A4), structure in the maps is 
 dominated by variations in the non-thermal spectral
index. 

Large variations in $\alpha$ can also be seen
in a direction away from the 
major axis,
indicating that there are real differences in $\alpha_{NT}$ between
the disk and halo.  The differences are in the sense of 
steeper (whiter on the maps) $\alpha_{NT}$ away from the plane.
Duric, Irwin \& Bloemen (1998), in a sensitive radio
continuum study of NGC~5775, have suggested that such a gradient
 is due to the superposition of a steep spectral index
halo and a flatter spectral index disk.
 There is also some evidence for steeper spectral indices 
in positions near discrete vertical extensions.
For example, a steep spectral index is seen in NGC~1421 just at the
base of a vertical extension on the west side of the galaxy
(at 3$^{\rm h}$~40$^{\rm m}$~7$\rasec$5, -13$\adeg$~38$\amin$~30$\asec$).
NGC~3044 shows strong steepening of the spectral index on
either side of the radio continuum peak where vertical extensions are
visible (see 
9$^{\rm h}$~51$^{\rm m}$~8$^{\rm s}$, 01$\adeg$~49$\amin$~15$\asec$ and
9$^{\rm h}$~51$^{\rm m}$~5$^{\rm s}$, 01$\adeg$~48$\amin$~25$\asec$).
Similarly, NGC~3556 and
NGC~5775 both show steepening of the spectral index away
from the major axis, especially in regions where 
extensions occur.  The exception appears to be 
the Seyfert/AGN galaxy, NGC~4388, which displays
a flat spectral index in the region of the nuclear outflow.  If 
flatter spectral indices occur in regions of active winds
(see also Fig. 2c of \cite{dib98}), then
steeper spectral indices near large discrete vertical features may be
typical of older features where the active winds have
ceased.

\subsubsection{Energy and Magnetic Field Maps}
\label{enmagmaps}

Using the minimum energy assumption, it is also possible to obtain
maps of magnetic field 
strength, B$_{min}$, and
CR 
energy density, u${_{min}^{CR}}$, averaged over a line of
sight.  These are computed (see \cite{pac70}; also see \cite{dur91}),
in cgs units, from:
$${\rm B}_{min}\,=\, {[6\,\pi\,(1+k)\,c_{12}\,V^{-1}\,L]}^{2/7}$$
$${\rm u}{_{min}^{CR}}\,=\, (1+k)\,c_{12}\,V^{-1}\,
{\rm B}_{min}^{-3/2}\,L$$
where $k$ is the heavy particle to electron ratio,  $V$ is the volume,
$L$ is the luminosity, and $c_{12}$ is one of ``Pacholczyk's constants"
which, for the convention S$_\nu\,\propto\,\nu^{\alpha}$, is determined from:
$$c_{12}\,=\,1.06 \times 10^{12}
\,\Bigl({{2\alpha+2}\over {2\alpha + 1}}\Bigr)
\,{{\Bigl[\nu{_1^{(1+2\alpha)/2}}\,-\,\nu{_2^{(1+2\alpha)/2}}\Bigr]}
\over
{(\nu{_1^{1+\alpha}}\,-\,\nu{_2^{1+\alpha}})}}
$$
Here $\nu_1$, $\nu_2$ are the lower and upper frequency cut-offs and
$\alpha$ is the spectral index.

We adopt the parameters, $k$ = 40, $\nu_1$ = 10$^7$ Hz (the conventional value),
and $\nu_2$ = 10$^{11}$ Hz.  Since the luminosity, volume, and
spectral index all vary across the galaxy, we compute the above quantities
pixel-by-pixel, taking $V\,=\,r^2\,l$, where $r$ is the length of a pixel
side
and $l$ is the line of sight distance through the galaxy at that pixel.
Each map therefore required 3 input maps:
the natural weighting 20 cm flux density map from which
the luminosity is computed, the matching resolution
spectral index map, and
a map of line of sight distances, $l$.  The latter was generated by assuming a 
spheroidal model galaxy which is inclined
by the upper limit inclination (Table~\ref{galcharacter})
with semi-major
axis, a, set to match the observed value, and an intrinsic axial ratio,
c/a, adjusted so that the model minor/major axis ratio 
(Table~\ref{radaxes}) also
 matched the observations.  The resulting u${_{min}^{CR}}$ 
and B$_{min}$ maps are
shown (major axis horizontal) in Appendix A.  
Mean, minimum and maximum values from the u${_{min}^{CR}}$ and
B$_{min}$ maps are 
presented in Table~\ref{enmag}.  Again, using the guidelines outlined
in \nocite{con92} Condon (1992), we find that a thermal
contribution should change 
u${_{min}^{CR}}$ and B$_{min}$ by less than 10\% 

Choosing a thin disk model to compute line of sight distances,
changing the inclination to the lowest possible value (cf.
Table~\ref{galcharacter}), adoping a lower upper-frequency cutoff,
or a different galaxy centre also each result in errors of order
10\%.  Differences between the 2 weightings of the data (i.e.
different beam sizes) can be 30\%, and adopting a different
heavy particle to electron energy ratio, $k$ (which can range from
1 to 100 in the literature) can produce an uncertainty of order a factor
of 2.  Thus the final values should be correct to within
factors of 2 - 3.  However,
the maps should illustrate the spatial variation in these values to a
higher degree of accuracy
(i.e. to
$\ltabouteq$ 50\%), since the same values of $k$ and upper 
frequency cutoff are used for each point.

There is some evidence that when the in-disk energy and magnetic
field strengths are higher (whiter on the maps), there is a greater
probability that the radio continuum emission will be more extended in
regions on either side of this emission.  For example, NGC~1421, NGC~3044,
and NGC~5775 all show broader (in the z direction)
 radio continuum emission on either side
of brighter in-disk regions.  On the other hand, a distinct radio spur in
NGC~1421 is above a region of weak energy in the disk and the
aforementioned correlations are not particularly obvious in either NGC~3556 or
NGC~4388.  Thus, there appears to be only a weak correlation, if any, between 
site of high energy and magnetic field and the presence of high latitude
radio continuum emission.

\subsubsection{Cosmic Ray Lifetimes and Propagation}
\label{crprop}

Table~\ref{enmag} also presents estimates of CR age,
$\tau_{CR}$, Alfv\'en
velocity, v$_A$, (\cite{pac70})
 and diffusion length, which gives the distance 
a cosmic ray particle
 could travel before decaying (from $\tau_{CR}$v$_A$).
The latter two quantities vary with the 
 ISM density which is unknown and will vary within and
between these galaxies.
We adopt
 a typical Galactic value of
0.18 cm$^{-3}$ which provides an estimate only of these quantities.
Nevertheless, it is clear that the diffusion lengths (Column 13), 
 which range from 1.4 to 4.9 kpc, are
 systematically less than the maximum vertical extent of the
observed extra-planar
radio emission  (corrected for beam smoothing), i.e.
8 kpc for NGC~1421, 8 kpc for NGC~3044, 10 kpc for NGC~3556,
11 kpc for NGC~4388 (if real, see \S\ref{rad-discrete}),
 and 18 kpc for NGC~5775.
For the diffusion length to match the observed distribution, the
mean ISM density would have to be as low as 0.002 to 0.04 cm$^{-3}$
which is unlikely to be the case over most of the ISM in
 these galaxies.  This
suggests that a) the high latitude radio continuum features
exist because CRs diffuse through low density ``chimneys"
(\cite{ni89}) or intercloud regions (see \cite{dib98}), and/or
b) an additional mechanism is required to move the CRs to high latitude,
examples being CR ``winds" (see \cite{blo91} and references therein),
Parker instabilities (\cite{ka96}), or
ram pressure stripping.  A similar
conclusion was reached for NGC~3432 by \nocite{enir97} English \&
Irwin (1997).

\section{Discussion}
\label{discussion}

The 16 edge-on galaxies in this sample
are all star forming galaxies which represent
a variety of star formation rates,
type, degree of interaction and
nuclear activity. 
 Yet all of them, except NGC~4517, show some evidence
for extended radio continuum emission away from the plane
apparently originating in the disk.  This 
suggests that high latitude radio continuum emission is common in
star forming disk galaxies, in contrast to the results of 
\nocite{hbd91} Hummel et al. (1991) (see \S\ref{sample}).  The difference
is probably due to our choice of longer wavelength (20 cm rather than 6 cm),
and target galaxies of larger angular size (up to 10.5$\arcmin$
rather than $<$ 4$\arcmin$; see comments on resolution below).

In Table~\ref{rank} we list the radio ranking, from
\S\ref{radio-rank}, where galaxies are ranked out
of 6 if there were additional D array data for them and out of 4 if not.
We take the error bar on the ranking to be the width of the bin.
 In addition, we list the star formation
rate (SFR$_{0.1\to100 M_{\odot}}$) and
the supernova energy input rate per unit star forming area
(\.E$\rm_A $) both from Table~\ref{luminosity} and listed in order
of decreasing \.E$\rm_A $.  We also list
the galaxy type (including Seyfert designation, if relevant),
 and information as to whether the galaxy appears to
be isolated, have companions, is in a group or cluster, or is
interacting.  The latter information is taken from the NASA/IPAC
Extragalactic Database (NED) 
\footnote{\small The NASA/IPAC Extragalactic Database (NED) is operated
by the Jet Propulsion Laboratory, California Institute of Technology,
under contract with the National Aeronautics and Space Administration.}
and the references, notes, and catalogues therein.
Before searching for correlations between the radio rank and other
properties of the sample, however, we must first consider how
S/N ratio and resolution might affect the results.

We computed a S/N ratio for each galaxy in two different ways.
First, for each array configuration from which the radio rank was
determined, we took
the peak galaxy intensity divided by the noise in the map 
 and then averaged these ratios to acquire a S/N for each object.
  Second, we found the ratio of total flux
(Table~\ref{luminosity}) to noise
(Table~\ref{beams}). The first method assumes that high latitude
radio emission
may scale with a local SFR (which is measured by the radio continuum
brightness) and the second method assumes that 
high latitude radio emission may scale
with a global SFR (as measured by the radio continuum flux).  By the
first method, the galaxies with the lowest S/N are NGC~5297, NGC~4517,
and NGC~4746.  
By the second method, the galaxies with the lowest
S/N are NGC~4388, NGC~3221, and NGC~5297.  If our results depend
strongly on the S/N ratio (and if the radio emission scales as we have
assumed)  then these low sensitivity galaxies  should
have the lowest radio rankings.  
From Table~\ref{rank}, however, other than NGC~4517, these galaxies
do show significant evidence for radio halos.  Galaxies such as
NGC5433, NGC~3735, and IC~0564, for example, have significantly 
higher S/N (by both measures) than these, yet lower radio
rankings.  Even NGC~4517, which shows no high latitude radio
emission, has a higher S/N (by the first method) than NGC~5297
(which does have high latitude emission) and a significantly higher S/N 
(by the second method) than NGC~4388, NGC~3221, and NGC~5297 (each
of which has high latitude emission).
Thus, while S/N must matter at some level, the radio rankings of Table~\ref{rank} do 
appear to be reflecting real
 differences between radio halos in these galaxies.

As for resolution,
all the galaxies are reasonably well-resolved along their minor axes.
For the C array uniformly weighted data, for example, 
there are between 2.7 and
9 beams across the minor axis (\S~\ref{measdim}).
The galaxies which have the fewest beams per minor axis are
IC~0564, NGC~3735,
and NGC~5433 and these do indeed have the lowest radio rankings
(other than NGC~4517).  Thus the probability of detecting extended
radio emission does appear to improve with spatial resolution.
One might expect that
lower resolution observations
would be more likely to detect radio halos because halos are 
considered to be broad-scale and smooth.
However, we found
(\S\ref{modeldisk}) that more discrete features appear as the resolution
improves and that discrete features can sometimes
be seen on the largest scales.  This suggests that
radio ``halos" consist, at least in part and possibly entirely
of discrete features combined with low spatial resolution.  Timescale
arguments are consistent with this hypothesis.  We have shown 
(\S\ref{crprop}) 
that general diffusion of CRs in the vertical direction 
over the entire star forming disk, such as would
produce a smooth halo, cannot account for the high latitude radio
continuum emission observed.
Other than the 3 low-resolution galaxies listed above, all other
 galaxies
in the sample have between 4 and 9 beams per minor axis.  An
 examination of 
the discrete features which are well-resolved suggests that 4 beams
per minor axis  appear to provide sufficient resolution to detect 
such features, if they exist. 
In the following sections, therefore, when we search for correlations
between the radio ranking of the galaxies and other properties,
we will do so for all the galaxies first, and then excluding the
3 low-resolution
galaxies: IC~0564, NGC~3735, and NGC~5433.

Finally, when searching specifically
for correlations with star formation indicators (\S~\ref{radio-sfr},
\S~\ref{radio-E})
the AGN in the sample could be excluded because of the
possible non-starburst
 nature of their high latitude emission.
There are two Seyfert galaxies in the sample (NGC~4388 and NGC~3735),
but only one of them (NGC~4388) shows evidence for a radio AGN and does
not fall on the FIR-radio continuum correlation (see Fig.~\ref{FIRradio}).
NGC~3735 does fall on the FIR-radio relation, 
suggesting that its nuclear activity
is a minor contributor to the global radio flux.
In addition, while the star forming galaxy, NGC~3432, does
not have an AGN, it is peculiar in that it
 does not fall on the FIR-radio relation.
Thus, for correlations with star formation indicators, we will
first consider all galaxies and then exclude 
both NGC~4388 and NGC~3432 from the
discussion due to their non-conformity to the FIR-radio relation.

``Radio halos", below,
is meant to mean any extra-planar radio continuum emission, as measured
by the radio ranking listed in Table~\ref{rank}. 

\subsection{ Radio Halos as a Function of Global Star Formation Rate}
\label{radio-sfr}

A comparison of the
SFR  
with the radio ranking 
(Table~\ref{rank}), shows no obvious correlation.
 For example, of the 5
starburst galaxies (SFR$_{0.1\to 100 M{_\odot}}$ greater than 
the M~82 value of
9 $\rm M_{\odot}\ yr^{-1}$, see also
 Fig.~\ref{FIRradio}) only one (NGC~5775)
has very strong extra-planar radio emission.
Galaxies with very low star formation 
rates (NGC~2683 and NGC~3432), on the other hand,
can have significant extra-planar
radio emission.
With 5 galaxies excluded as explained above
(3 because of low resolution and 2 because of non-compliance to
the FIR-radio continuum relation),
the conclusion is the same.  To put this more quantitatively,
we computed the linear correlation coefficient between SFR and radio ranking,
with the radio ranking weighted (as to whether the observations are
ranked out of 4 or 6).  The correlation coefficient for all 16 galaxies
is -0.29 and, for the subset of 11 galaxies, it is 0.007, in either case,
significantly less (in absolute value) than 1.  Thus, there is
no correlation between radio rank and global SFR.

\subsection{Radio Halos as a Function of \.E$\rm_A $}
\label{radio-E}

From a sample of 9 edge-on galaxies,
Dahlem et al. (1995) found some evidence to suggest that it is
the energy input rate per unit star forming area, \.E$\rm_A $,
rather than the global SFR,
 which determines whether
or not radio halos will be present.

Again, 
a comparison of \.E$\rm_A $
with the radio rank (Table~\ref{rank}) shows no clear
correlation.
Galaxies with strong radio halos, for example, occur
over a wide range of \.E$\rm_A $ (cf. NGC~5775, NGC~3556, and
NGC~3432).  For the complete set, the weighted correlation
coefficient (as described above, but using \.E$\rm_A $
rather than SFR) is 0.18. Taking into account
 errors
of up to 50\% on \.E$\rm_A $ (see \S~\ref{derive}) in various
ways (e.g. random, systematic) cannot increase the coefficient by
more than $\sim$ 20\%.

Excluding the 5 galaxies listed above, the correlation coefficient
increases to 0.42. (Again, this cannot increase significantly, when
errors in \.E$\rm_A $ are taken into account.)  The probability
that a correlation coefficient this high or higher could occur
from a random sample of the same size is 20\%.  Thus, there could
be a weak correlation between these two properties for the subset
of galaxies, but it is far from convincing.  
 
The one galaxy, however, with the lowest value of
\.E$\rm_A $ (i.e. NGC~4517)  shows no evidence at all for
 extra-planar radio emission.
Dahlem et al. have suggested that there exists a lower cutoff to
 \.E$\rm_A $ below which ``blowout"
cannot occur, and therefore a radio halo will not form.
 Our data suggest that, if such a threshold exists,
it would be at 
 $\approx$ 0.04 $\times$ 10$^{-3}$ erg s$^{-1}$ cm$^{-2}$, i.e.
the level observed for NGC~4517.
  Correcting for the different methods of computing $\nu_{SN}$
(see \S~\ref{derive}),
this value does agree with the lower threshold 
of 10$^{-4}$ erg s$^{-1}$ cm$^{-2}$
given by Dahlem et al. (1995). 

Thus, our results are consistent with the existence of a lower 
threshold to
\.E$\rm_A $ for the formation of a radio halo but, at best, allows
for only a weak correlation between
\.E$\rm_A $ 
and the strength of the radio halo as we have defined it amongst
a subset of the galaxies.

\subsection{Radio Halos as a Function of Galaxy Type}
\label{radio-type}

We find no relation between the presence of high latitude radio continuum
and Hubble type, or the presence or absence of bars, although
the latter may be rather difficult to classify, given the 
edge-on orientation of the galaxies.  
With the 3 low-resolution galaxies
excluded,
again there is  no correlation.
The two Seyfert galaxies, NGC~3735
and NGC~4388 also show very different levels of extended radio emission.

\subsection{Radio Halos as a Function of Galaxy Environment}
\label{radio-env}

This issue is more difficult to assess since it is not always clear whether
the presence of a companion implies an interaction. It
is now recognized that even when an interaction is in progress, a starburst
may or may not be occurring depending on the details and timing of
the event (\cite{mh96}). 
Table~\ref{rank} indicates that most of the galaxies in the sample
are not isolated.  Only three galaxies, NGC~3044, NGC~3556, and NGC~1421
might arguably be considered isolated.  Yet all three show strong
evidence for high latitude radio continuum emission and two of these
are known to have high latitude HI shells (NGC~3044: \cite{li97};
NGC~3556: \cite{ki97}).  Thus, it would appear
that interactions are not necessary
to produce radio halos.  

It has been suggested, however, that interactions
may enhance radio halos due to gravitational perturbations
for cases in which the SFR is relatively low
(e.g. NGC~4631, Dahlem et al. 1995; NGC~3432, English \& Irwin 1997).
In our sample, the
galaxies with the lowest values of
\.E$\rm_A $ are NGC~3432, NGC~2613, NGC~5297, and NGC~4517. Of these,
the first 3 are either interacting or in a group and also show 
evidence for radio halos, NGC~3432 having the strongest.
NGC~4517 does not have a radio halo but is paired with UGC~07685
and could also be in a small group.  Thus, while the data do not
rule out this possibility, they do not strongly support it either.
Nothing more definitive can be said if the 3 low-resolution galaxies
are exluded.
It will likely take a larger sample with better understood interaction
properties
to adequately investigate this possibility.

Another way in which environment may play a role in forming extra-planar
features is through
ram pressure stripping.  Virgo Cluster galaxies, such as NGC~4388
(see \S\ref{rad-discrete}), 
 may fall into
this category (see \cite{cbvk90}).

\subsection{Radio Halos as a Relic of Past Activity}

For the 5 galaxies for which we could compute radio properties
(Table~\ref{enmag}), mean CR lifetimes 
range from 3 $\times$ 10$^7$ yr to 1.3 $\times$ 10$^8$ yr.  Using the
lower magnetic field strengths which occur in the halo regions, 
this range increases
 to  6 $\times$ 10$^7$ yr to 2 $\times$ 10$^8$ yr.
Thus, in general, the lifetime of halo CRs in these galaxies is
$\approx$ 10$^8$ yr.

The lifetime of any given OB association is of order a few $\times$
10$^7$ yr.  Thus, if cosmic rays escape from the disk over specific
star forming regions, the resulting discrete feature may remain
visible even after the region below it has ceased to be actively
star-forming.
 While star formation can take place globally
over longer timescales, this effect could still mask or weaken a correlation
between radio halos and SFR or 
between halos and \.E$\rm_A $, if either correlation existed initially.
The effect should be strongest for starbursts.
Theoretical models of starbursts in  mergers
suggest starburst durations of $\approx$ 5 $\times$ 10$^7$ yr
(\cite{mh96}).  Recent Infrared Space Observatory observations of
luminous IR galaxies are also suggesting relatively short
starburst durations of 1 - 2 $\times$ 10$^7$ yr (\cite{lutz96}). 
Thus if a radio halo has resulted from starburst activity, it
could remain even after all evidence for the starburst has disappeared.   
  This could explain why we do not see a clear correlation between radio
halos and
\.E$\rm_A $ (\S\ref{radio-E}).  A search
for correlations between discrete flat spectral index features (e.g.
a signature of active winds)
and in-disk star forming regions, or between shorter-lived high
latitude features (such as extra-planar H$\alpha$) and in-disk
star forming regions may be more fruitful.

\section{Conclusions}
\label{conclusions}

We have observed 16 edge-on galaxies in the radio continuum using the
VLA at
20 cm in its C configuration and have supplemented this with 6 and/or
20 cm data in D configuration for a subset of the galaxies. 
The range of spatial scales probed is 11$\asec$ to either 4$\arcmin$
(C array 20 cm data) or 15$\arcmin$ (D array 20 cm data)
compared to the optical dimensions of the galaxies which range from
1.6$\arcmin$ to 10.5$\arcmin$.  The 
sample was chosen by high inclination, 
 by angular size, and by radio flux density and
represents a wide range of star formation rates (only 6 galaxies
are starbursts), Hubble type, and degree of isolation or interaction. 
The purpose of the survey was to search for high latitude 
($\gtabouteq$ 1 kpc) radio continuum
emission. 

Of these 16 edge-on galaxies, we find that
all but one galaxy (NGC~4517) show evidence
for such emission. Of these, 8 are new
detections.  Thus, extra-planar radio emission
 appears to be common in star forming
galaxies.  The radio emission is
seen on a variety of  scales  and in a variety of ways, including
large
radio axial ratios (b/a), the presence of discrete features extending
away from the major axis, or broad scale emission extending beyond
the projection of a modeled thin radio disk.  Discrete features
can be seen on a variety of spatial scales including the broadest scales,
suggesting that 
radio halos may, in part and perhaps completely, 
be discrete features as seen with large beams.
Indeed, high resolution appears to be necessary to detect high
latitude features.
 Nuclear outflow is seen in the case of the Seyfert/AGN NGC~4388 and radio
spurs seen in NGC~3044 could originate
from its nucleus.  However, even in these galaxies
 there is evidence for
some disk features and discrete features appear to originate in the
disk in all other cases.  

For 5 galaxies, we could form maps of
spectral index, CR energy density and
magnetic field strength.  The emission is predominantly non-thermal
and variations in the maps are significantly 
greater than variations expected from
a thermal contribution.  The spectral index shows steepening away from
the plane in regions of discrete extensions but 
there is no strong correlation 
between high latitude radio features and the energy or
magnetic field strength. Diffusion alone appears to be
 insufficient to transport CRs to high latitude
for reasonable mean ISM densities.
Thus, they could be transported through low density regions and/or
assisted by other mechanisms such as winds.

All galaxies, with the exception of the Seyfert/AGN, NGC~4388, and the low
SFR galaxy, NGC~3432 (see \cite{enir97}) fall on the FIR - radio continuum
correlation.  In addition, we have extended the IR surface brightness --
colour relation found for starburst galaxies by Lehnert \& Heckman~(1996b)
to the lower SFR galaxies of this sample.

We have ranked the
galaxies into broad categories, depending on the strength and
extent of the extra-planar radio emission over all detectable scales,
and have searched for correlations between radio rank and other properties,
including global SFR, supernova energy input rate per unit
area, \.E$\rm_A$, galaxy environment, and
galaxy type. We find no clear correlation with the latter two properties.
 It is difficult to assess the
correlation between radio rank and galactic environment, however, since 
the degree to which a galaxy may
be interacting is not always clear.
Thus a larger sample is likely needed.  Nevetheless, since 3
apparently isolated galaxies do have strong radio halos,
it would appear that interactions are not necessary 
to generate high latitude radio emission. 

We find no clear correlations with either SFR or \.E$\rm_A$.
If we exclude 5 galaxies because of low resolution or  
non-compliance with the FIR-radio continuum relation, then the
data allow for a weak correlation with 
\.E$\rm_A$ (correlation coefficient of 0.4), but the result is not
 convincing. 
The one galaxy with no evidence for high latitude radio emission 
(NGC~4517) does, however, have a value of
\.E$\rm_A$ which agrees with the lower threshold given
by Dahlem et al. 1995. 
 The lack of a good correlation between
radio halos and \.E$\rm_A$  might be
explained by the difference in timescales.
With lifetimes of $\sim$ 10$^8$ yr, radio halos may represent the integration
of star formation activity over that timescale, whereas quantities such
as SFR and \.E$\rm_A$ represent (in comparison) snapshots of activity
which might last for 
$\sim$ 10$^7$ yr in specific star forming regions or
$\ltabouteq$ 5 $\times$ 10$^7$ yr for starbursts.  Thus, if a radio halo --
starburst correlation originally exists, it may weaken with time.
This suggests that it may be better to search for correlations
between \.E$\rm_A$ and extra-planar H$\alpha$ emission or to search
for correlations between discrete flat-spectral index features (such as
would indicate the presence of active winds) and in-disk star
forming regions in order to search for a starburst origin for the high
latitude features.
If high latitude radio continuum emission does result from
star forming activity, then the existence of a radio halo
 could provide a useful signature that a 
galaxy has indeed experienced a starburst in the past.

\noindent {\bf Acknowledgements:}  We gratefully acknowledge the
assistance of Michael Earl in preparing the figures for publication.
The Digitized Sky Surveys were produced
at the Space Telescope Science Institute under U.S. Government grant
NAG W-2166.  The images of these surveys are based on photographic data
obtained using teh Oschin Schmidt Telescope on Palomar Mountain and
the UK Schmidt Telescope.  The plates were processed into the present
compressed digital form with the permission of these institutions.
We acknowledge the use of NASA's {\em SkyView} facility 
(http://skyview.gsfc.nasa.gov) located at NASA Goddard Space Flight
Center.  This research has made use of the NASA/IPAC Extragalactic Database
(NED) which is operated by the Jet Propulsion Laboratory, California
Institute of Technology, under contract with the National Aeronautics and
Space Administration.  We are grateful to the Australia National
Telescope Facility for the use of their Karma software package.


      \nocite{RC3}
      \nocite{tul88}
      \begin{table}
      \dummytable\label{galcharacter}
      \end{table}

\begin{table}
\dummytable\label{beams}
\end{table} 

      \begin{table}
      \dummytable\label{radaxes}
      \end{table}

\begin{table}
\dummytable\label{luminosity}
\nocite{RC3}
\nocite{lang74}
\nocite{sanmir96}
\nocite{sbns89}
\nocite{iraspt}
\end{table}

\begin{table}
\dummytable\label{enmag}
\nocite{dur91}
\nocite{pac70}
\end{table}

\begin{table}
\dummytable\label{rank}
\nocite{old86}
\nocite{irw94}
\nocite{sch86}
\nocite{sha90}
\nocite{cha80}
\nocite{giu90}
\nocite{tur88}
\nocite{sur93}
\nocite{zar97}

\end{table}

\newpage
\appendix{
\centerline {Appendix A: Maps of
16 Edge-On Spiral Galaxies.} 

For individual galaxies,  1-2 pages are 
devoted to total intensity
maps (synthesized beam is shown at bottom left of each) and derived
spectral index and energy and magnetic field maps, if the latter are
available.
\footnote{\small The plots were made using
Karma Visualization software developed by the CSIRO Australia
Telescope National Facility.}  The maps are as follows: 

\begin{enumerate}
\item Contours of uniformly weighted 20-cm C-array data
overlayed on DSS.  A logarithmic display
of the B passband images is used so that the full optical extent 
of each galaxy is apparent. (This is the largest image 
and is at the top or lefthand side of the page.)
Contour levels begin at 2$\sigma$ and are listed in Table~A1.
\item Contours (plus greyscale)
 of the naturally weighted 20-cm C-array
data.  The optical axes are shown and 
have arrow tips if they extend beyond the contour map.  Optical
data are taken from the RC3, or from NED if the latter values
differ significantly from the RC3.
Contour levels, beginning at 2$\sigma$,
are listed in Table~A1.
\item Contours (plus greyscale) of the naturally weighted 20-cm D-array
data with optical axes shown.
 Contour levels,
starting at 2$\sigma$, are listed in Table~A2.
\item Contours (plus greyscale) of the naturally weighted 6-cm D-array
data. 
Contour levels, starting at 2$\sigma$, are listed in Table~A3.
\item Contours (plus greyscale) of spectral index data, where
S$_\nu\,\propto\,\nu^{\alpha}$ (\S~\ref{specindex}).
  White represents steep values of
$\alpha$ while black represents flat  $\alpha$.
Contour levels are listed in Table~A4.
\item  Logarithmic greyscale 
of minimum energy in CRs (\S~\ref{enmagmaps}) with
white representing high energy and black representing low energy.
 20-cm C-array 
naturally weighted contours are
overlaid.  Contour levels and 
greyscale ranges are listed in Table~A5. 
\item  Logarithmic 
greyscale of magnetic field maps (\S~\ref{enmagmaps})
with 20-cm C-array (naturally weighted) contours
overlaid.  
Contour levels and greyscale ranges are listed in Table~A5.  White 
represents high magnetic fields
and black represents low.
\end{enumerate} 
}


\newpage
\figcaption[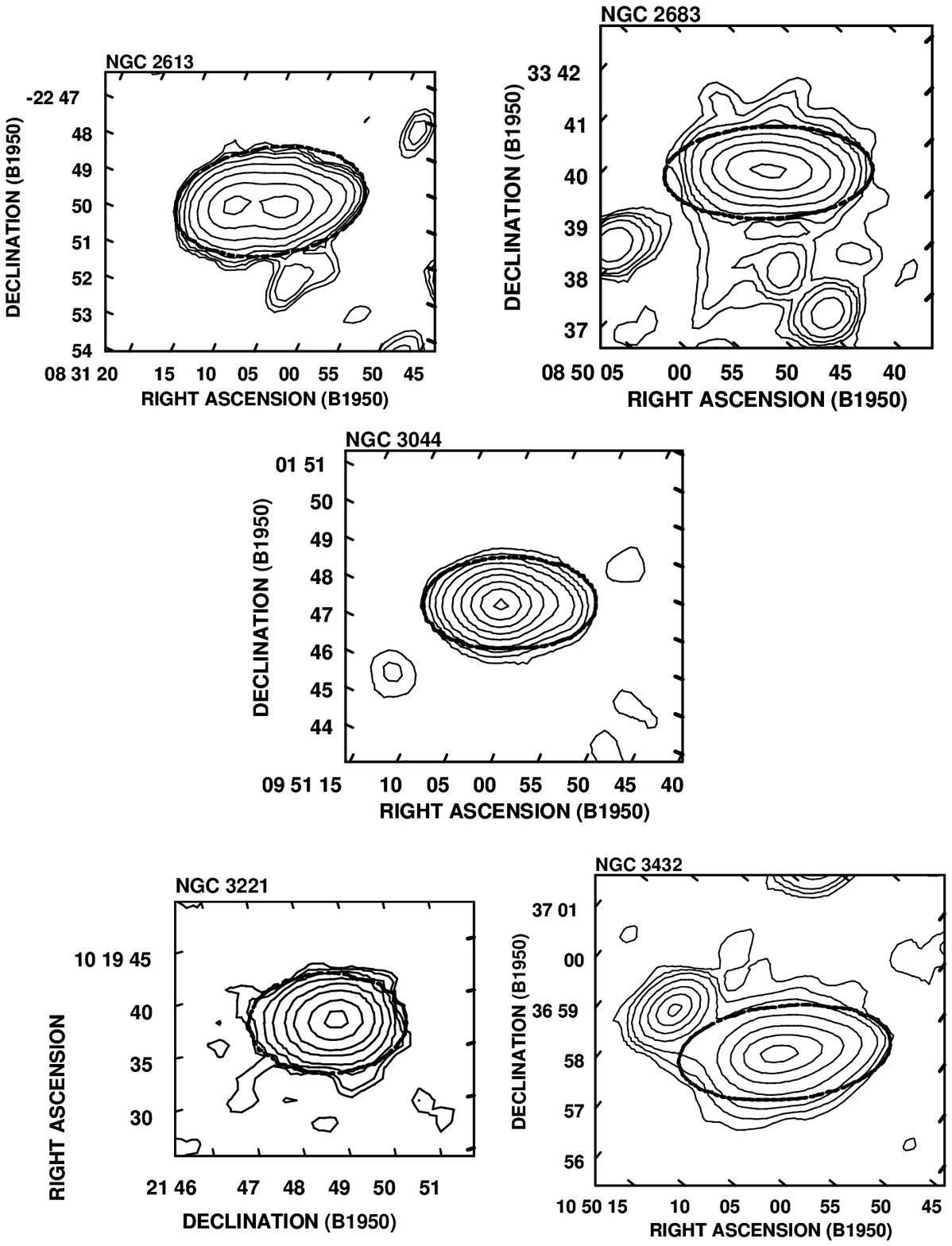, 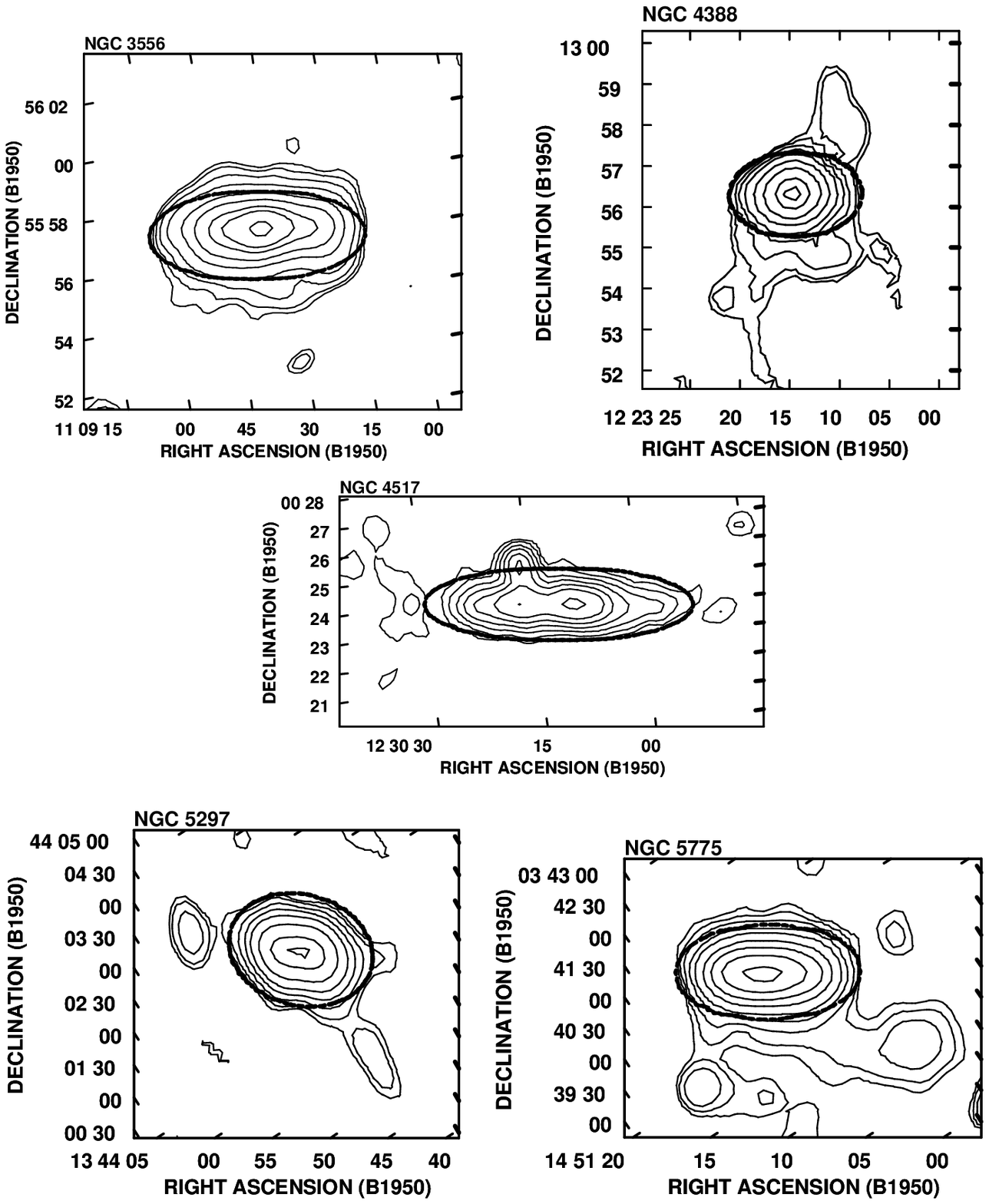]{Model thin disks are super-imposed
as a thick dashed line 
on 20~cm D Array naturally weighted data (\S~\ref{modeldisk}).
 For each image, the major axis has been rotated to
align with the x axis and the first contour is at 2$\sigma$.
See Appendix A for other, non-rotated displays of these data. 
\label{thindisk}}

\figcaption[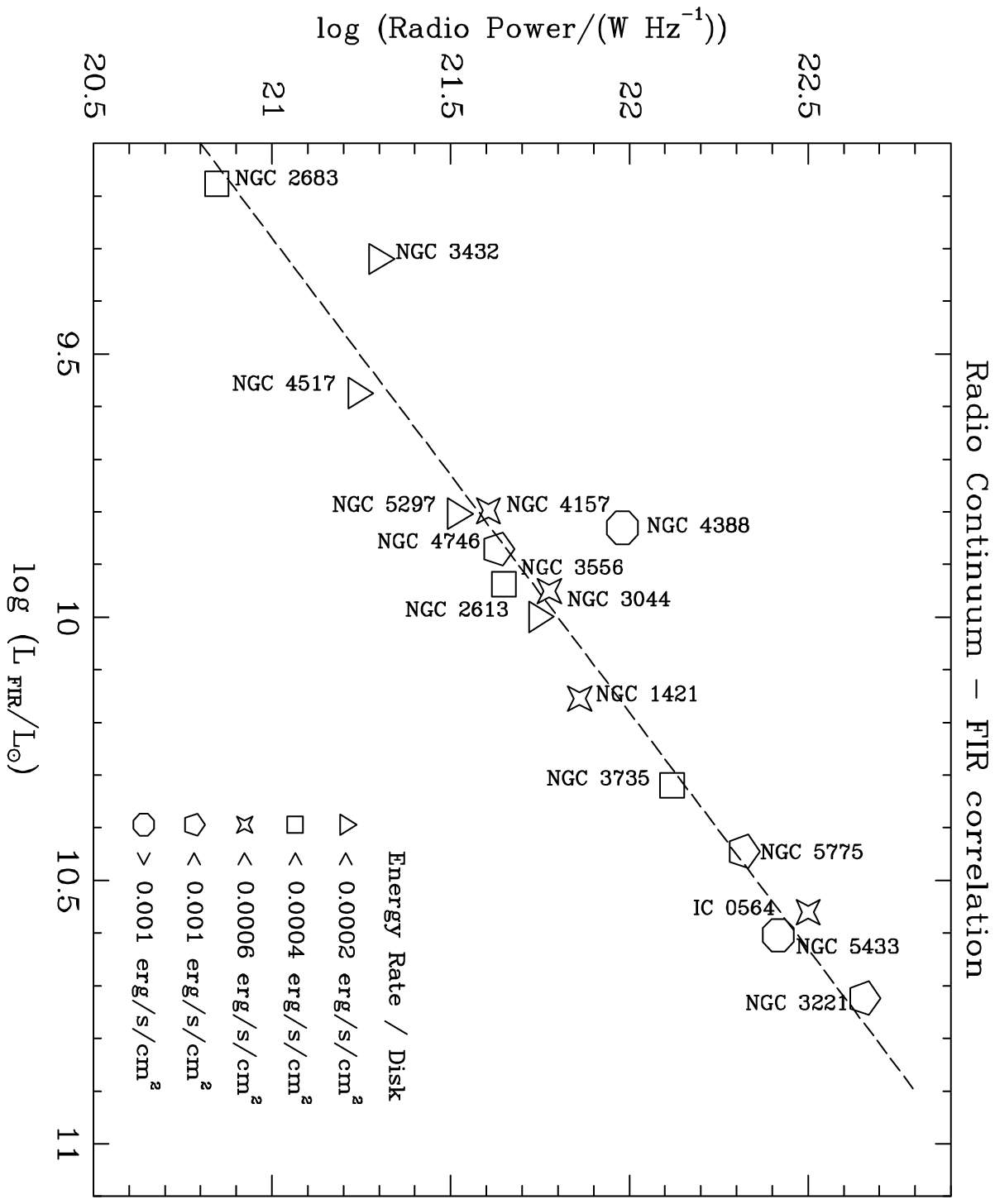]
{Radio Continuum -- Far Infrared Correlation. 
The markers indicate supernova energy input rate per unit
star forming area (cf. Table~\ref{luminosity}). 
\label{FIRradio}}

\figcaption[fig3.ps]
{An estimate of the supernova energy input rate 
per unit star forming area is 
plotted against the dust temperature (FIR colour).  
See \S~\ref{colour-luminosity}.
\label{dustenrate}}


\begin{thebibliography}{}
 
\bibitem[{Beichman} { et~al.} 1988]{iraspt}
{Beichman, C.~A.}, {Neugebauer, G.}, {Habing, H.~J.}, {Clegg, P.~E.}, \&
  {Chester, T.~J.}, editors 1988,
\newblock { Infrared Astronomical Satellite Point Source Catalog},
\newblock NASA (RP-1190)
 
\bibitem[{Bloemen} 1991]{blo91}
{Bloemen, H.} 1991,
\newblock in { The Interpretation of Modern Synthesis Observations of Spiral
  Galaxies}, edited by {Duric, N.} \& {Crane, P.~C.}, volume~18, page~27,
  A.S.P. Conf. Ser
 
\bibitem[{Bloemen} { et~al.} 1993]{bdi93}
{Bloemen, H.}, {Duric, N.}, \& {Irwin, J.~A.} 1993,
\newblock in { Proc 23rd International Cosmic Ray Conference}, edited by
  {Leahy, D.~A.}, {Hicks, R.~B.}, \& {Venkatesan, D.}, volume~2, page 279,
  World Scientific
 
\bibitem[{Bregman} 1980]{bre80}
{Bregman, J.~N.} 1980,
\newblock \aap, 236, 577
 
\bibitem[{Cayatte} { et~al.} 1990]{cbvk90}
{Cayatte, V.}, {Balkowski, C.}, {van Gorkom, J.~H.}, \& {Kotanyi, C.} 1990,
\newblock \aj, 100, 604
 
\bibitem[{Chamaraux} { et~al.} 1980]{cha80}
{Chamaraux, P.}, {Balkowski, C.}, \& {G/'erard, E.} 1980,
\newblock \aap, 83, 38
 
\bibitem[{Colbert} { et~al.} 1996]{col96}
{Colbert, E. J.~M.}, {Baum, S.~A.}, {Gallimore, J.~F.}, {O'Dea, C.~P.}, \&
  {Christensen, J.~A.} 1996,
\newblock \apj, 467, 551
 
\bibitem[{Condon} 1987]{con87}
{Condon, J.~J.} 1987,
\newblock \apjs, 65, 485
 
\bibitem[{Condon} 1992]{con92}
{Condon, J.~J.} 1992,
\newblock \araa, 30, 575
 
\bibitem[{Condon} { et~al.} 1990]{chss90}
{Condon, J.~J.}, {Helou, G.}, {Sanders, D.~B.}, \& {Soifer, B.~T.} 1990,
\newblock \apjs, 73, 359
 
\bibitem[{Dahlem} 1997]{dahl97}
{Dahlem, M.} 1997,
\newblock \pasp, 109, 1298
 
\bibitem[{Dahlem} { et~al.} 1995]{dlg95}
{Dahlem, M.}, {Lisenfeld, U.}, \& {Golla, G.} 1995,
\newblock \apj, 444, 119
 
\bibitem[{Dahlem} { et~al.} 1997]{dah97}
{Dahlem, M.}, {Petr, M.~G.}, {Lehnert, M.}, {Heckman, T.~M.}, \& {Ehle, M.}
  1997,
\newblock \aap, 320, 731
 
\bibitem[{de~Vaucouleurs} \& {Corwin} 1976]{RC2}
{de~Vaucouleurs, G. de~Vaucouleurs, A.} \& {Corwin, H. G.~J.} 1976,
\newblock { Second Reference Catalogue of Bright Galaxies (RC2)},
\newblock University of Texas Press, New York
 
\bibitem[{de~Vaucouleurs} { et~al.} 1991]{RC3}
{de~Vaucouleurs, G.}, {de~Vaucouleurs, A.}, {Corwin, H.~G.}, {Buta, R.~J.},
  {Paturel, G.}, \& {P., F.} 1991,
\newblock { Third Reference Catalogue of Bright Galaxies (RC3)},
\newblock Springer-Verlag New York
 
\bibitem[{Dettmar} 1992]{det92}
{Dettmar, R.-J.} 1992,
\newblock Fundamentals of Cosmic Physics, 15, 143
 
\bibitem[{Dressel} \& {Condon} 1978]{drecon78}
{Dressel, L.~L.} \& {Condon, J.~J.} 1978,
\newblock \apjs, 36, 53
 
\bibitem[{Duric} 1991]{dur91}
{Duric, N.} 1991,
\newblock in { The Interpretation of Modern Synthesis Observations of Spiral
  Galaxies}, edited by {Duric, N.} \& {Crane, P.~C.}, volume~18, page~17,
  A.S.P. Conf. Ser
 
\bibitem[{Duric} { et~al.} 1998]{dib98}
{Duric, N.}, {Irwin, J.~A.}, \& {Bloemen, H.} 1998,
\newblock \aap, 331, 428
 
\bibitem[{English} \& {Irwin} 1997]{enir97}
{English, J.} \& {Irwin, J.~A.} 1997,
\newblock \aj, 113, 2006
 
\bibitem[{Franco} { et~al.} 1991]{fr91}
{Franco, J.}, {Ferrini, F.}, {Ferrara, A.}, \& {Barsella, B.} 1991,
\newblock \apj, 366, 443
 
\bibitem[{Giuricin} { et~al.} 1990]{giu90}
{Giuricin, G.}, {Bertotti, G.}, {Mardirossian, F.}, \& {Mezzetti, M.} 1990,
\newblock \mnras, 247, 444
 
\bibitem[{Golla} \& {Hummel} 1994]{gh94}
{Golla, G.} \& {Hummel, E.} 1994,
\newblock \aap, 284, 777
 
\bibitem[{Heckman} { et~al.} 1990]{ham90}
{Heckman, T.~M.}, {Armus, L.}, \& {Miley, G.~K.} 1990,
\newblock \apjs, 74, 833
  
\bibitem[{Helou} { et~al.} 1985]{helsr85}
{Helou, B.}, {Soifer, B.~T.}, \& {Rowan-Robinson, M.} 1985,
\newblock \apjl, 298, L7
 
\bibitem[{Howk} \& {Savage} 1997]{hs97}
{Howk, J.~C.} \& {Savage, B.~D.} 1997,
\newblock \aj, 114, 2463
 
\bibitem[{Hummel} 1980]{hum80}
{Hummel, E.} 1980,
\newblock \aaps, 41, 151
 
\bibitem[{Hummel} { et~al.} 1991]{hbd91}
{Hummel, E.}, {Beck, R.}, \& {Dettmar, R.-J.} 1991,
\newblock \aaps, 87, 309
 
\bibitem[{Hummel} { et~al.} 1985]{hphd85}
{Hummel, E.}, {Pedlar, A.}, {van~der Hulst, J.~M.}, \& {Davies, R.~D.} 1985,
\newblock \aaps, 60, 293
 
\bibitem[{Hummel} \& {Saikia} 1991]{hs91}
{Hummel, E.} \& {Saikia, D.~J.} 1991,
\newblock \aa, 249, 43
 
\bibitem[{Hummel} \& {van~der Hulst} 1989]{hvh89}
{Hummel, E.} \& {van~der Hulst, J.~M.} 1989,
\newblock \aaps, 81, 51
 
\bibitem[{Hummel} { et~al.} 1983]{hvk83}
{Hummel, E.}, {van Gorkom, J.}, \& {Kotanyi, C.} 1983,
\newblock \apj, 267, L5
 
\bibitem[{Hunter} { et~al.} 1986]{hggrl86}
{Hunter, D.~A.}, {Gillet, F.~C.}, {Gallagher, J.~S.}, {Rice, W.~L.}, \& {Low,
  F.~J.} 1986,
\newblock \apj, 303, 171
 
\bibitem[{Irwin} \& {Sofue} 1996]{is96}
{Irwin, J.} \& {Sofue, Y.} 1996,
\newblock \apj, 464, 738
 
\bibitem[{Irwin} 1991]{irw91}
{Irwin, J.~A.} 1991,
\newblock in { The Interpretation of Modern Synthesis Observations of Spiral
  Galaxies}, edited by {Duric, N.} \& {Crane, P.~C.}, volume~18, page~59, San
  Francisco: A.S.P.
 
\bibitem[{Irwin} 1994]{irw94}
{Irwin, J.~A.} 1994,
\newblock \apj, 429, 618
 
\bibitem[{Irwin} 1995]{irw95}
{Irwin, J.~A.} 1995,
\newblock \pasp, 107, 715
 
\bibitem[{Kamaya} { et~al.} 1996]{ka96}
{Kamaya, H.}, {Mineshige, S.}, {Shibata, K.}, \& {Matsumoto, R.} 1996,
\newblock \apj, 458, L25
 
\bibitem[{King} \& {Irwin} 1997]{ki97}
{King, D.~L.} \& {Irwin, J.~A.} 1997,
\newblock New Astronomy, 2, 251
 
\bibitem[{Lang} 1974]{lang74}
{Lang, K.~R.} 1974,
\newblock { Astrophysical Formulae}, page 558,
\newblock New York: Springer-Verlag
 
\bibitem[{Lee} \& {Irwin} 1997]{li97}
{Lee, S.-W.} \& {Irwin, J.~A.} 1997,
\newblock \apj, 490, 247
 
\bibitem[{Lehnert} \& {Heckman} 1996a]{lh96s}
{Lehnert, M.~D.} \& {Heckman, T.~M.} 1996a,
\newblock \apj, 462, 651
 
\bibitem[{Lehnert} \& {Heckman} 1996b]{lh96}
{Lehnert, M.~D.} \& {Heckman, T.~M.} 1996b,
\newblock \apj, 472, 546

\bibitem[{Lutz} { et~al.} 1996]{lutz96}
{Lutz, D.}, {Genzel, R.}, {Sternberg, A.}, {Netzer, H.}, {Kunze, D.},
  {Rigopoulou, D.}, {Sturm, E.}, {Egami, E.}, {Feuchtgruber, H.}, {Moorwood, A.
  F.~M.}, \& {de~Graauw, T.} 1996,
\newblock \aap, 315, L137
 
\bibitem[{Mihos} \& {Hernquist} 1996]{mh96}
{Mihos, J.~C.} \& {Hernquist, L.} 1996,
\newblock \apj, 464, 641
 
\bibitem[{Niklas} { et~al.} 1997]{nik97}
{Niklas, S.}, {Klein, U.}, \& {Wielebinski, R.} 1997,
\newblock \aap, 322, 19
 
\bibitem[{Norman} \& {Ikeuchi} 1989]{ni89}
{Norman, C.~A.} \& {Ikeuchi, S.} 1989,
\newblock \apj, 345, 372
 
\bibitem[{Oldenwald} 1986]{old86}
{Oldenwald, S.~F.} 1986,
\newblock \apj, 310, 86
 
\bibitem[{Pacholczyk} 1970]{pac70}
{Pacholczyk, A.~G.} 1970,
\newblock { Radio Astrophysics: Non-thermal Processes in Galactic and
  Extragalactic Sources},
\newblock San Francisco: W. H. Freeman \& Co.

\bibitem[{Phillips} \& {Malin} 1982]{pmalin82} 
{Phillips, M.~M.} \& {Malin, D.~F.} 1982, 
\newblock MNRAS 199, 905
 
\bibitem[{Sanders} \& {Mirabel} 1996]{sanmir96}
{Sanders, D.~B.} \& {Mirabel, I.~F.} 1996,
\newblock \araa, 34, 749
 
\bibitem[{Schneider} { et~al.} 1986]{sch86}
{Schneider, S.~E.}, {Helou, G.}, {Salpeter, E.~E.}, \& {Terzian, Y.} 1986,
\newblock \aj, 92, 742
 
\bibitem[{Shapiro} \& {Field} 1976]{sf76}
{Shapiro, P.~R.} \& {Field, G.~B.} 1976,
\newblock \apj, 205, 762
 
\bibitem[{Sharp} 1990]{sha90}
{Sharp, N.~A.} 1990,
\newblock P.A.S.P., 102, 109
 
\bibitem[{Soifer} { et~al.} 1989]{sbns89}
{Soifer, B.~T.}, {Boehmer, L.}, {Neugebauer, G.}, \& {Sanders, D.~B.} 1989,
\newblock \aj, 98, 766
 
\bibitem[{Sorathia} 1994]{sor94}
{Sorathia, B.} 1994,
\newblock A Radio Continuum Survey of Edge-on Spiral Galaxies,
\newblock Master's thesis, Queen's University
 
\bibitem[{Stone} { et~al.} 1988]{sto88}
{Stone, J.~L., J.}, {Wilson, A.~S.}, \& {Ward, M.~J.} 1988,
\newblock \apj, 330, 105
 
\bibitem[{Surace} { et~al.} 1993]{sur93}
{Surace, J.~A.}, {Mazzarella, J.}, {Soifer, B.~T.}, \& {Wehrle, A.~E.} 1993,
\newblock \aj, 105, 864
 
\bibitem[{Tully} 1988]{tul88}
{Tully, R.~B.} 1988,
\newblock { Nearby Galaxies Catalog},
\newblock Cambridge Univ. Press
 
\bibitem[{Turner} { et~al.} 1988]{tur88}
{Turner, K.~C.}, {Helou, G.}, \& {Terzian, Y.} 1988,
\newblock \pasp, 452, 457
 
\bibitem[{Vogler} \& {Pietsch} 1996]{vp96}
{Vogler, A.} \& {Pietsch, W.} 1996,
\newblock \aap, 311, 35
 
\bibitem[{V{\"o}lk} 1991]{vol91}
{V{\"o}lk, H.~J.} 1991,
\newblock in { The Interstellar Disk-Halo Connection in Galaxies}, edited by
  {Bloemen, H.}, volume 144, page 345, I.A.U. Symp.
 
\bibitem[{Wehrle} 1987]{weh87}
{Wehrle, A.~E.} 1987,
\newblock { The Radio Structure of Active Nuclei in Edge-on Spiral and Seyfert
  Galaxies},
\newblock PhD thesis, California Univ., Los Angeles
 
\bibitem[{Zaritsky} { et~al.} 1997]{zar97}
{Zaritsky, D.}, {Smith, R.}, {Frenk, C.}, \& {White, S. D.~M.} 1997,
\newblock \apj, 478, 39
 
\end{thebibliography}
\end{document}


\newcommand\cola {\null}
\newcommand\colb {&}
\newcommand\colc {&}
\newcommand\colg {&}
\newcommand\cold {&}
\newcommand\cole {&}
\newcommand\colf {&}
\newcommand\coli {&}
\newcommand\colh {&}
\newcommand\colj {&}
\newcommand\eol{\\}
\newcommand\extline{&&&&&&&&&\eol}

\begin{deluxetable}{lccccccccc}
\ptlandscape
\tablewidth{0pc}
\tablecaption{Fluxes, Luminosities and Star Formation Rate Indicators. \label{luminosity}}
\tablecomments{Flux densities (2), (3) are measured from D array data (D)
or C array data (C), see \S~4.2.
 Error bars are the larger of
a) $\sigma$$\sqrt N$, where $\sigma$ is the rms map noise and
 N is the number of beams occupied by the galaxy, or
b) the measured flux in an equivalent region of the background
or c) the variation in flux due to choice of different measurement
regions.  
The radio power (4)
is computed from P$_{20cm}$ = 4 $\pi$ D$^2$ S$_{20cm}$, with D from Column (3),
Table~1.
The optical luminosity (5) is derived from 
the total face-on apparent B magnitude (B$\rm^O_T$) given
in the \nocite{RC3}RC3, using 5.41 for the absolute blue magnitude of
the sun (\nocite{lang74}Lang 1974).
Columns (6) and (7) list
 the IR and FIR luminosities, following 
\nocite{sanmir96}
Sanders \& Mirabel~(1996).  
  Flux densities  from
 \nocite{sbns89}Soifer~et.~al.(1989) were used.  Those not in
Soifer~et~al. were taken from the \nocite{iraspt}
IRAS Point Source Catalog~(1988) and are marked with a 
* in column (1).
Ratios in 
column (8) also used these catalogues. 
The star formation rate
(9) and supernova input energy rate per unit star forming area (10)
are described in \S~4.2. {\it a)} IR Values for IC~0564 are taken
from \nocite{sur93} Surace et al. (1993)
who resolved IC~0564 from its nearby companion, IC~0563.
 }
\tablehead{
\cola Name
\colb S$_{6cm}$
\colc S$_{20cm}$
\colg log P$_{20 cm}$
\cold log $\rm (L_{B}/L_\odot)$
\cole log $\rm (L_{IR}/L_\odot)$
\colf log $\rm (L_{FIR}/L_\odot)$
\coli S60/S100
\colh SFR$_{0.1\to 100 M_{\odot}}$  
\colj \.E$\rm_A$\eol
\cola 
\colb mJy 
\colc mJy
\colg log Watts Hz$^{-1}$
\cold  \cole \colf 
\coli
\colh $\cal M_{\odot} \rm  \ yr^{-1}$ \colj $\rm 10^{-3} \ erg \ s^{-1} \ cm^{2}$\eol
\cola (1) \colb (2) \colc (3) \colg (4) \cold (5) \cole (6) \colf (7)  
\coli (8) 
\colh (9)  \colj(10)\eol
}
\startdata
\cola  NGC~1421\colb 37 $\pm$ 7 \colc 89 $\pm$ 2 (C)
\colg  21.86
\cold  10.57\cole 10.44\colf  10.16
\coli 0.46\colh 7.21\colj 0.48\eol
\cola  NGC~2613 * \colb  \colc 69 $\pm$ 4 (C)
\colg  21.74
 \cold  11.01 \cole 10.29\colf 10.00 \coli  0.33\colh  5.13\colj 0.10\eol
\cola NGC~2683 \colb \colc 84 $\pm$ 4 (D)
\colg  20.85
\cold  10.16\cole 9.47\colf   9.18\coli 0.30\colh 0.766\colj 0.24 \eol
\cola  IC~0564$^a$ \colb \colc 37 $\pm$ 2 (C)
\colg  22.50
 \cold  10.79\cole 11.00\colf  10.56\coli 0.38\colh 25.8\colj 0.56\eol
\cola NGC~3044 \colb 42 $\pm$ 7 \colc 106 $\pm$ 3 (D)
\colg  21.78
\cold  10.36\cole 10.22\colf   9.95\coli 0.50\colh 4.27\colj 0.54 \eol
\cola NGC~3221 \colb \colc 109 $\pm$ 2 (D)
\colg  22.66
\cold  10.62\cole 11.00\colf  10.72\coli 0.38\colh 26.0\colj 0.88\eol
\cola NGC~3432 \colb \colc 120 $\pm$ 3 (D)
\colg  21.30
\cold   9.89\cole 9.53\colf   9.32\coli 0.52\colh 0.886\colj 0.14\eol
\cola  NGC~3556 \colb 89 $\pm$ 7 \colc 279 $\pm$ 5 (C)
\colg  21.65
\cold  10.35\cole 10.21\colf   9.94\coli 0.42\colh 4.26\colj 0.30\eol
\cola  NGC~3735 \colb \colc 80 $\pm$ 4 (C)
\colg  22.12
\cold  10.70\cole 10.62\colf  10.32\coli 0.38\colh 10.8\colj 0.30\eol
\cola  NGC~4157 \colb \colc 202 $\pm$ 5 (C)
\colg  21.61
 \cold 9.69\cole  10.09\colf    9.80\coli 0.35\colh  3.18\colj 0.52\eol
\cola NGC~4388 \colb 73 $\pm$ 7 \colc 200 $\pm$ 10 (D) 
\colg  21.98
\cold  10.43\cole 10.22\colf  9.83\coli 0.58\colh 4.40\colj 1.08 \eol
\cola NGC~4517 * \colb \colc 38 $\pm$ 4 (D)
\colg  21.24
\cold  10.74\cole 9.83\colf   9.57\coli 0.27\colh 1.76\colj 0.04\eol
\cola NGC~4746 * \colb \colc 46 $\pm$ 4 (C)
\colg  21.63
\cold  10.13\cole 10.15\colf   9.87\coli 0.39\colh 3.65\colj 0.62\eol
\cola NGC~5297 * \colb \colc 23 $\pm$ 5 (D)
\colg  21.52
\cold  10.67\cole 10.12\colf   9.80\coli 0.27\colh 3.46\colj 0.18\eol
\cola  NGC~5433 \colb \colc 59 $\pm$ 4 (C)
\colg  22.42
\cold  10.72\cole 10.87\colf  10.60\coli 0.56\colh 19.3\colj 1.56\eol
\cola NGC~5775 \colb 76 $\pm$ 11 \colc 275 $\pm$ 5 (D)
\colg  22.32
\cold  10.47\cole 10.72\colf  10.44\coli 0.46\colh 13.7\colj 0.70\eol
\enddata
\end{deluxetable}


\newcommand\cola {\null}
\newcommand\colb {&}
\newcommand\colc {&}
\newcommand\cold {&}
\newcommand\cole {&}
\newcommand\colf {&}
\newcommand\colg {&}
\newcommand\colh {&}
\newcommand\coli {&}
\newcommand\colj {&}
\newcommand\coll {&}
\newcommand\colm {&}
\newcommand\coln {&}
\newcommand\eol{\\}
\newcommand\extline{&&&&&&&&&\eol}

\begin{deluxetable}{lcccccccccccc}
\ptlandscape
\tablewidth{0pc}
\tablecaption{Radio Properties\label{enmag}}
\tablecomments{ Columns (2) to (4) give the mean, minimum, and maximum
spectral index measured from the 
natural weighting spectral index maps (Appendix A). Values in parentheses
are measured from the uniform weighting maps (not shown). 
Columns (5) to (7) give the cosmic ray energy
density and columns (8) to (10) give
the magnetic field strength
(mean, minimum and maximum
from the natural weighting maps shown in Appendix A),
assuming minimum energy.  The CR lifetime, 
$\tau_{CR}\,\propto\,$B$_{min}^{-3/2}$ (Column 11), and 
Alfv\'en velocity, v$_A$ = B$_{min}$/$\sqrt{4\pi\rho}$ (Column 12), are
computed from the mean B$_{min}$ (Column 8).
 v$_A$ is an order of
magnitude estimate assuming a constant mean ISM density of 0.18 cm$^{-3}$
for each galaxy. The estimate of diffusion length (Column 13) is 
computed from $\tau_{CR}$v$_A$.  (See \nocite{pac70}
Pacholczyk 1970,  \nocite{dur91}
Duric 1991, and \S~4.3.2 for more details.)  
 }             
\tablehead{
\cola Name 
\colb \multicolumn{3}{c}{$\alpha$}\cold
\multicolumn{3}{c}{$\rm u_{min}^{CR}$}\colg
\multicolumn{3}{c}{$\rm B_{min}$}
\colj $\tau_{CR}$
\colm v$_A$
\coln Diffusion \eol
\cola \colb \colc \cold 
\cole 
\colf $10^{-12}$ erg cm$^{-3}$
\colg 
\colh 
\coli $\mu$Gauss
\colj 
\coll  $10^6$ yr
\colm km s$^{-1}$
\coln Distance\eol
\cola 
\colb mean
\colc min
\cold max
\cole mean
\colf min
\colg max
\colh mean
\coli min
\colj max
\coll 
\colm 
\coln kpc \eol
\cola (1) \colb (2)  \colc (3) \cold (4) \cole (5) \colf (6) 
\colg (7) \colh (8) \coli (9) \colj (10)
\coll (11) \colm (12) \coln (13) \eol 
}
%
\startdata

\cola NGC~1421
\colb  -0.80 (-0.66)
\colc  -1.48 (-1.28)
\cold -0.11 (0.21) 
\cole    6.53
\colf     3.72
\colg    49.3
\colh    6.8\coli    5.2\colj   19.0
\coll     70\colm     35\coln 2.5 \eol
\cola NGC~3044
\colb  -0.60 (-0.81)
\colc  -1.04 (-1.38)
\cold  0.22 (-0.13)
\cole    4.77
\colf     1.66\colg    10.7\colh    5.8\coli    3.5\colj    8.9
\coll     46\colm     30\coln 1.4\eol
\cola NGC~3556
\colb -1.03 (-0.95)
\colc -1.58 (-1.41)
 \cold -0.56 (-0.47)
\cole     6.76\colf     4.00\colg     15.5\colh    7.0\coli    5.4\colj   10.7
\coll    132\colm     36\coln 4.9\eol
\cola NGC~4388
\colb -0.78 (-0.76)
\colc  -1.52 (-1.15)
\cold 0.00 (-0.29)
\cole   18.4\colf    9.24\colg   33.1\colh    11.6\coli    8.3\colj   15.6
\coll     30\colm     60\coln 1.8\eol
\cola NGC~5775
\colb -0.93 (-0.97)
\colc -1.81 (-1.92)
\cold -0.18 (-0.14)
\cole    7.99\colf     4.13\colg    19.4\colh    7.6\coli    5.5\colj   11.9
\coll     89\colm     39\coln 3.6\eol
\enddata
\end{deluxetable}